\documentstyle[amscd,12pt,psamsfonts,leqno,amssymb]{amsart}
\input xypic

\newcommand{\F}{{\mathcal{F}}}
\newcommand{\R}{{\mathcal{R}}_N}
\newcommand{\M}{{\mathcal{M}}_N}
\newcommand{\hr}{\tilde{\mathcal{R}_N}}
\newcommand{\hm}{\tilde{\mathcal{M}_N}}
\newcommand{\Lm}{{\mathcal{L}}_m}
\newcommand{\X}{X_N}
\newcommand{\gr}{{\operatorname{deg}}\ }
\newcommand{\poin}{{\mathcal{P}}}
\newcommand{\dua}{\hat{X}}

\theoremstyle{plain}
\newtheorem{teo}{Theorem}[section]
\newtheorem{cor}[teo]{Corollary}

\newtheorem{prop}[teo]{Proposition}

\theoremstyle{remark}
\newtheorem{rem}{Remark}[section]
\newtheorem{demo}{Proof}[section]

\begin{document}

\date{}
\title[]{Higher order addition laws on Abelian varieties  and the
fractional quantum Hall effect}
\author{Juan Mateos Guilarte${\;}^{(1)}$ \\ Jos\'e Mar\'{\i}a Mu\~{n}oz Porras${\;}^{(2)}$}
\thanks{${\;}^{(1)}$ Departamento de F\'{\i}sica, Ingenier\'{\i}a y Radiolog\'{\i}a
M\'edica. Facultad de Ciencias. Universidad de Salamanca. Salamanca 37008. SPAIN\\
\indent ${\;}^{(2)}$ Departamento de Matem\'{a}ticas. Facultad de
Ciencias. Universidad de Salamanca. Salamanca 37008. SPAIN}

\maketitle

 \begin{abstract}
Addition formulas for theta functions of arbitrary order are shown
and applied to the theoretical understanding of the fractional
quantum Hall effect in a multi-layer two-dimensional many-electron
system under periodic conditions.
 \end{abstract}

\section{Introduction}
As is well known, classical addition formulas for theta functions
are formulas of degree two. In this paper we prove addition
formulas of arbitrary degree for theta functions. The explicit
generalized addition formulas are stated in 2.5 and 2.6 and the
main ingredients in the proof are the cube theorem and the isogeny
theorem of Mumford (\cite {1} and \cite {4} ).

We apply these results to offer a formulation of the fractional
quantum Hall effect in a multi-layer many-electron system and
possible generalizations. In the study of the ordinary quantum
Hall effect under periodic conditions (\cite {5}), the basic
geometric object is an algebraic torus $$\Sigma=\frac{{\mathbf
C}}{{\mathbf Z}+\tau{\mathbf Z}},$$ and study of the isogeny
$$\xymatrix{\varphi_N\colon \Sigma^N\ar[r] &   \Sigma^N}$$
$$\varphi(x_1,\ldots,x_N)=(x_1+\cdots+x_N,x_1-x_2,\ldots,x_1-x_N)$$
determines the change of coordinates from one-particle to
center-of-mass and relative coordinates in the space of wave
functions of the system. The change from one-particle coordinates,
$(x_1,\ldots,x_N)$, to center-of-mass and relative coordinates,
$(x_1+\cdots+x_N,x_1-x_2,\ldots,x_1-x_N)$ is crucial in the
definition of the Laughlin wave function describing the ground
state of the FQHE on the complex plane. On the algebraic torus,
$\Sigma$, this change of coordinates becomes the isogeny mentioned
above. In the second quantization formalism, to extend the
Laughlin variational principle for the fractional quantum Hall
effect to the periodic case thus requires the use of generalized
addition formulas for elliptic theta functions, see \cite {5}.

The goal of this work is generalize this construction to an
 arbitrary algebraic torus of dimension g $X=\frac{{\mathbf C}}{{\mathbf Z}^g+\tau{\mathbf Z}^g}$
 ( where $\tau=(\tau_{ij}$ is a complex $g\times g$
 symmetric
 matrix with the imaginary part positive definite). From a physical
 point of view, this means that we are considering a system of many
 electrons moving on g two-dimensional layers with different
 periodic conditions defined by the period matrix $\tau_{ij}$.
 As in our earlier paper \cite {5} , we are naturally lead to study
 the isogenies
$$\xymatrix{\varphi_N\colon X^N\ar[r] &   X^N}$$
$$\varphi(x_1,\ldots,x_N)=(x_1+\cdots+x_N,x_1-x_2,\ldots,x_1-x_N).$$
The behaviour of the global sections of some line bundles over
$X^N$ under the isogeny $\varphi_N$ will determine the vector
space of wave functions of the system of electrons under the
generalized periodic conditions . This behaviour and the link
between wave functions depending either on $(x_1,...,x_N)$ or
$(x_1+...+x_N,x_1-x_2,...,x_1-x_N)$ is completely described by the
generalized addition formulas. We thus obtain a complete
description of the space of wave functions of the FQH effect in a
multi-layer two-dimensional electron system. In fact, in the real
physical situation it suffices to consider a diagonal period
matrix $\tau$, unless tunnel effects between different layers are
taken into account, see ( \cite{14} ).

Having done this , we also consider the Fourier-Mukai transform of
some line bundles over $X^N$ determined by the generalized
Haldane-Rezayi wave functions and the semi-stability of these
transforms. We show how the slope of these bundles is related to
the Hall conductivity, which therefore appears as a topological
invariant. There is an isomorphism between the Fourier-Mukai
transforms for any number of electrons N and, thus, the Hall
conductivity depends only on the center-of-mass dynamics
characterized by the Haldane-Rezayi states.

\noindent The organization of the paper is as follows: In Section
\S.2, we show the main theorems and establish the generalized
addition formulas for Abelian varieties. In Section \S.3, the
vector spaces for the quantum ground states are constructed in
terms of higher order odd theta functions. Section \S.4 is devoted
to studing the Fourier-Mukai transform of the line bundles over
$X^N$ related to the quantum vector spaces. In Section \S.5, all
these developments are applied to the analysis of the fractional
quantum Hall effect in multi-layer periodic electron systems.
Finally, in Section \S.6 a comparison with the physics literature
is offered and some obscure points are clarified.

\section{Generalized addition formulas for abelian varieties}\label{s:1}
Let $X$ be an abelian variety of dimension $g$ over the field
${\mathbb{C}}$ of complex numbers. Let us define the following
family of morphisms: $$\xymatrix{M,m_{ij},s_{ij}\colon X
\times\overset{N}\cdots\times X  \ar[r]    &   X}$$
\begin{align} M(x_1,\ldots,x_N)&=x_1+\ldots+x_N     \notag \\
                m_{ij}(x_1,\ldots,x_N)&=x_i+x_j     \notag \\
                s_{ij}(x_1,\ldots,x_N)&=x_i-x_j     \notag \end{align}

$$\xymatrix {p_i \colon  X \times \ldots \times X \ar[r]& X} $$
will be the natural projections.
\begin{teo} {\it Generalized Cube Theorem}
\label{t:Generalizedcubetheorem}

For any symmetric invertible sheaf $L$ on $X$ one has a natural
isomorphism: $$M^*L\simeq
\left(\bigotimes_{i<j}m_{ij}^*L\right)\otimes\left(\bigotimes_{i=1}^{N}
p_i^*L^{\otimes -N+2}\right)$$
\end{teo}

\begin{demo}
This follows from the cube theorem, \cite {1}, and induction over
$N$.
\end{demo}

\begin{cor}\label{c:Generalizedcubetheorem}
For any symmetric invertible sheaf $L$ over $X$, one has a natural
isomorphism:
$$M^*L\otimes\left(\bigotimes_{i<j}s_{ij}^*L\right)\simeq
p_1^*L^{\otimes N}\otimes\cdots\otimes p_N^*L^{\otimes N}$$
\end{cor}

\begin{demo}
By the Theorem~\ref{t:Generalizedcubetheorem} one has:
$$M^*L\otimes\left(\bigotimes_{i<j}s_{ij}^*L\right)\simeq
\bigotimes_{i<j}\left(m_{ij}^*L \otimes s_{ij}^*L\right) \otimes
\left(\bigotimes_{i=1}^{N} p_i^*L^{\otimes -N+2}\right)$$

Let us denote by $\xymatrix{p_{ij}\colon X \times
\overset{N}\cdots\times X \ar[r]    &   X\times X}$ the projection
on the factors $(i,j)$ and by $\xymatrix{\pi\colon X\times X
\ar[r]    &   X}\ (i=1,2)$ the natural projections. One has: $$
m_{ij}^*L \otimes s_{ij}^*L\simeq
p_{ij}^*\left(\xi^*\left(\pi_1^*L\otimes
\pi_2^*L\right)\right)\simeq p_{ij}^*\left(\pi_1^*L^{\otimes
2}\otimes\pi_2^*L^{\otimes2}\right)$$ where $\xymatrix{\xi\colon
X\times X \ar[r] &   X\times X}$ is the morphism:
$\xi(x,y)=(x+y,x-y)$

We therefore have:
    {\small $$
    \begin{aligned}
    M^*L\otimes\left(\bigotimes_{i<j}s_{ij}^*L\right)&\simeq
    \bigotimes_{i<j} p_{ij}^*\left(\pi_1^*L^{\otimes
    2}\otimes\pi_2^*L^{\otimes2}\right) \otimes
    \left(\bigotimes_{i=1}^{N}p_i^*L^{\otimes -N+2}\right) \notag \\
    &\simeq \bigotimes_{i<j}\left( p_i^*L^{\otimes 2}\otimes
    p_j^*L^{\otimes2}\right) \otimes
    \left(\bigotimes_{i=1}^{N}p_i^*L^{\otimes -N+2}\right) \simeq
    \bigotimes_{i=1}^{N}p_i^*L^{\otimes N} \notag \end{aligned}$$}

\end{demo}

Let us consider the morphism of Abelian varieties:
    {\small
    $$\begin{aligned}
    \xi_N \colon X\times\overset{N}\cdots\times X
    & \longrightarrow X\times\overset{r}\cdots\times X
    \qquad\qquad (r=\frac{N(N-1)}{2}+1)\notag
    \\
    (x_1, ...., x_n) &\longmapsto
    (x_1+...+x_N, x_1-x_2, ..., x_{N-1}-x_N )   \notag
    \end{aligned}$$}
By Corollary~\ref{c:Generalizedcubetheorem} one has an
isomorphism:
$$\xi_N^*\left(p_1^*L\otimes\cdots\otimes p_r^*L\right)\simeq
M^*L\otimes\left(\bigotimes_{i<j} s_{ij}^*L\right)\simeq
p_1^*L^{\otimes N}\otimes\cdots\otimes p_r^*L^{\otimes N}$$ which
induces a homomorphism between the vector spaces of global
sections:
    $$
    \xi_N^*\colon
    H^0(X,L)\otimes\overset{r}\cdots \otimes H^0(X,L)
    \,\longrightarrow\,
    H^0(X,L^{\otimes N})\otimes\overset{r}\cdots \otimes
    H^0(X,L^{\otimes N})
    $$

For applications to the study of the quantum Hall effect under
periodic conditions, it is very important to compute explicitly
the homomorphism $\xi_N^*$(see  \cite{5} and the last Section of
this paper).

Observe that the kernel of $\xi_N$ is $\Delta (\X)$, where $\X$ is
the $N-$torsion subgroup of $X$ and $\xymatrix{\Delta\colon
X\ar@{^{(}->}[r]    &   X\times\overset{N}\cdots \times X}$ is the
diagonal immersion.

The morphism $\xi_N$ factors as follows:
$$\xymatrix{Z=X\times\overset{N}\cdots \times X \ar[r]^{\phi_N}  &
Y=Z/\Delta(\X)   \ar@{^{(}->}[r]^{i}    &   X\times\cdots^r \times
X}$$ $$\xi_N=i\circ\phi_N$$

Let us set ${\mathcal{L}}=\left(p_1^*L\otimes\overset{r}\cdots
\otimes p_r^*L\right).$

One has that
$\phi^*_N{\mathcal{L}}_{|_{Y}}=M^*L\otimes\left(\bigotimes_{i<j}
s_{ij}^*L\right)=\M.$

We can now consider the morphism: $$\xymatrix{\varphi_N\colon
Z\ar[r] &   Z}$$
$$\varphi(x_1,\ldots,x_N)=(x_1+\cdots+x_N,x_1-x_2,\ldots,x_1-x_N)$$
and define an invertible sheaf $\R$ on $Z$ by:
$$\R=\left(p_1^*L\otimes\cdots \otimes p_N^*L\right)\otimes
\left(\otimes s_{ij}^*L\right)$$

One has a commutative diagram:

$$\xymatrix{Z\ar[r]^{\xi_N}\ar[dr]_{\varphi_N}  &
X\times\overset{r}\cdots \times X \ar[d]^{\pi_{1\ldots N}}   \\
                                                &   Z                                                   }$$
$\pi_{1\ldots N}$ being the projection on the $N$ first factors.
$\pi_{1\ldots N}$ induces an isomorphism $\xymatrix{Y\ar[r]^{\sim}
&   Z}$ such that: $$\pi_{1\ldots N}^*\R\simeq {{\mathcal{L}}}\
\text{ and }\ \xi_N^*{\mathcal{L}}=\varphi_N^*\R\simeq\M.$$

But $\varphi_N$ is an isogeny of kernel $\Delta(\X)$ and the
problem of computing the homomorphism $\xi_N^*$ is reduced to
computing the homomorphism:
$$\xymatrix{\varphi_N^*:H^0(Z,\R)\ar[r]  &   H^0(Z,\M)}$$

To compute this homomorphism, we can apply the Mumford theory of
algebraic theta functions, \cite{2}.

To make explicit computations, let us fix a principal polarization
(p.p.) $\Theta$ on the abelian variety $X$, and assume that
$L={\mathcal{O}}_X\left(m\Theta\right)$; then, one has that
${\mathcal{M}}\simeq\bigotimes_{i=1}^{N}p_i^*{\mathcal{O}}_X(N.m\Theta).$

For any invertible sheaf $\F$ on $X$, let us denote by $K(\F)$ the
subgroup of $X$ which leaves $\F$ invariant under translations
($K(\F)={x\in X: T_x^*\F\simeq{\F}}$) and by ${\mathcal{G}}(\F)$
the theta-group of $\F$.

In our case, one has:

$$K(L)=X_m=\text{ subgroup of $m$-torsion points of }X$$
$$K(L^{\otimes N})=X_{N.m} \ \text{ and  }\ \ K(L)=N\cdot
K(L^{\otimes N})\subset X_{N.m}$$

The isomorphism $\varphi^*\R\simeq\M$ implies that: {\small
$$K(\M)=K(L^{\otimes N})\times\overset{N}\cdots \times
K(L^{\otimes N})=X_{Nm}\times\overset{N}\cdots \times
X_{Nm}\supset K(\R)\supset X_{m}\times\overset{N}\cdots \times
X_{m}$$}

For any invertible sheaf $L={\mathcal{O}}_X(D)$ on an Abelian
variety $X$ of dimension $g$, let us denote by $\gr(L)$ the number
$D^g$.

\begin{prop}\label{p:1.3}

$$|K(\R)|=N^{2(N-2)g}\cdot m^{2Ng}$$ $$\gr\R=(N g)!\
N^{(N-2)g}\cdot m^{Ng}$$

\end{prop}

\begin{demo}
Observe that $\mathrm{ker}\varphi_N=\Delta(\X)\simeq\X$. One then
has that $$\gr
\varphi_N^*\R=\gr\varphi_N^*\cdot\gr\R=N^{2g}\cdot\R$$ and $$\gr
\varphi_N^*\R=(Ng)!N^{Ng}m^{Ng}$$

Therefore: $\gr\R=(Ng)!\ N^{(N-2)g}m^{Ng}.$
\end{demo}

The structure of the group $K(\R)$ is given by the following
theorem:

\begin{teo}\label{t:1.4}
$K(\R)$ is the subgroup of points $\varphi_N(p)=(x_1+\cdots
+x_N,x_1-x_2,\ldots,x_1-x_N)\in X\times\overset{N}\cdots\times X$
such that: $p=(x_1,\ldots,x_N)\in X_{Nm}\times\overset{N}\cdots
\times X_{Nm}$ and $x_1+\cdots +x_N \in X_m$.

In particular, $K(\R)$ has subgroups isomorphic to
$X_{N}\times\overset{N}\cdots \times X_{N}$ given by:
$$\xymatrix{X_{m}\times\overset{N}\cdots \times
X_{m}\ar@{^{(}->}[r] &   K(\R)}$$
$$\xymatrix{(x_1,\ldots,x_N)\ar[r]  &   (x_1,\ldots,x_N)}$$ (with
respect to the natural immersion $X_m=N\cdot X_{Nm}\subset
X_{mN}$) and : $$\xymatrix{X_{N}\times\overset{N-2}\cdots \times
X_{N}\ar@{^{(}->}[r] &   K(\R)\ar@{^{(}->}[r]   &
X\times\overset{N}\cdots \times X}$$
$$\xymatrix{(x_2,\ldots,x_{N-1})\ar[r]  &
(0,-x_2,\ldots,-x_{N-1},x_2+\cdots +x_{N-1})    &   }$$
\end{teo}

\begin{demo}

Let $\hat{X}=\mathrm{Pic}^0(X)$ be the dual abelian variety. From
the exact sequence: $$\xymatrix@C=0.5cm{
  0 \ar[r] & \X \ar[rr]^{\Delta} && X\times\overset{N}\cdots \times X \ar[rr]^{\varphi_N} && X\times\overset{N}\cdots \times X \ar[r] & 0 }$$
one deduces the existence of the following dual exact sequence:
$$\xymatrix@C=0.5cm{
  0 \ar[r] & \hat{X}_N \ar[rr]^{\Delta} && \hat{X}\times\overset{N}\cdots \times \hat{X} \ar[rr]^{\varphi^*_N} && \hat{X}_N\times\overset{N}\cdots \times \hat{X}_N \ar[r] & 0 }$$
which means that given a point $p=(x_1,\ldots,x_N)\in K(\M)$, one
has: $$T^*_{\phi(p)}\R\otimes\R^{\otimes -1}\simeq
p_1^*M\otimes\cdots\otimes p_N^*M$$ for a certain invertible sheaf
$M$ of degree zero on $X$. By restricting this equality to
$X\times\{e\}\times\cdots\{e\}$ we compute $M$ and obtain the
following isomorphism: $$T^*_{\phi(p)}\R\otimes\R^{\otimes
-1}\simeq p_1^*\left(T^*_{x_1+\cdots+x_N}L\otimes L^{\otimes
-1}\right)\otimes\cdots\otimes
p_N^*\left(T^*_{x_1+\cdots+x_N}L\otimes L^{\otimes -1}\right)$$

Then, $T^*_{\phi(p)}\R\simeq \R$ if and only if $x_1+\cdots+x_N\in
X_m$. The rest of the theorem follows easily from this result.

\end{demo}

\begin{rem}
We have constructed two subgroups, $X_m\times\overset{N}\cdots
\times X_m$ and $X_N\times\overset{N-2}\cdots \times X_N$ of
$K(\R)$. Thus if $(m,N)=1$, a general element of $K(\R)$ has the
form:
$$(y_1,y_2-x_2,\ldots,y_{N-1}-x_{N-1},y_N+x_2+\cdots+x_{N_1})$$
where $(y_1,\ldots,y_N)\in X_m^N$ and $(x_2,\ldots,x_{N-1}\in
X_N^{N-2}).$
\end{rem}

Let us fix compatible theta-structures (\cite{2} )    on $L$ and
$L^{\otimes N}$. These theta-structures induce compatible
theta-structures on $\R$ and  $\M$ and decompositions:

$$\begin{aligned} &K(L)\simeq A(L)\times B(L) \ ,\ \ \ \ \
A(L)\simeq ({\mathbb{Z}}/{m{\mathbb{Z}}})^g    \notag \\
&K(L^{\otimes N}\simeq A(L^{\otimes N})\times B(L^{\otimes N}) \
,\ \ \ \ \
A(L^{\otimes N})\simeq ({\mathbb{Z}}/{mN{\mathbb{Z}}})^g    \notag \\
&K(\M)\simeq A(L^{\otimes N})^N\times B(L^{\otimes N})^N    \notag
\\ &K(\R)\simeq A(\R)\times B(\R)      \notag \end{aligned}$$

where $B(\R)\subset B(L^{\otimes N})^N  $, and by
Theorem~\ref{t:1.4} one has: $$B(L)^N\subset B(\R), \ \ \
mB(L^{\otimes N})^{N-2}\subset B(\R)$$

$$B(L)^N=N.B(L^{\otimes N})^N$$ in such a way that $B(\R)$ is the
subgroup of $B(L^{\otimes N})^N$ generated by $m.B(L^{\otimes
N})^{N-2}$ and $N.B(L^{\otimes N})^N$.

We have natural isomorphisms \cite{2}: $$\begin{aligned}
&H^0(X,L)=V_m=\{\text{functions} \xymatrix{B(L)\ar[r]   &
{\mathbb{C}}}\}   \notag \\ &H^0(X,L^{\otimes
N})=V_{Nm}=\{\text{functions} \xymatrix{B(L^{\otimes N})\ar[r]   &
{\mathbb{C}}}\}   \notag \end{aligned}$$

For each $d\in B(L)$, let $\delta_d$ be the global section of $L$
defined by the characteristic function of $d$, and for each $b\in
B(L^{\otimes N})$ let $\delta_b$ be the corresponding global
section of $L^{\otimes N}$.

Observe that $H^0(Z,\R)$ is a ${\mathbb{C}}$-vector space of
dimension $N^{(N-2)g}m^{Ng}$ and $H^0(Z,\M)$ is a
${\mathbb{C}}$-vector space of dimension $N^{Ng}m^{Ng}$. The
following result give us an explicit description of the
homomorphism: $$\xymatrix{\varphi^*_N\colon H^0(Z,\R)\ar[r] &
H^0(Z,\M)}$$

\begin{teo}\label{t:1.6}
Let us assume that $L$ and $L^{\otimes N}$ have compatible
theta-structures satisfying the above conditions. For each $d\in
B(\R)$ one has: $$\varphi^*_N(\delta_d)=\lambda\cdot
\sum_{\displaystyle b\in B(\M)\atop f(b)=d} \delta_d$$ where
$\lambda\in {\mathbb{C}}$ is a constant which we will assume to be
equal to 1.
\end{teo}

\begin{demo}
This follows from the isogeny theorem, \cite{1} and \cite{4}.
\end{demo}

This result allow us to give more explicit expressions for
$\varphi^*_N$.

Given $d=(d_1,\ldots,d_n)\in N\cdot
B(L)^N=[({\mathbb{Z}}/{m{\mathbb{Z}}})^g ]^N\subset B(\R)$, let us
denote by $\delta_d$ the element:
$$\delta_d=\delta_{d_1}\otimes\cdots\otimes\delta_{d_N}(_{\otimes
\atop {i>j \atop j\geq 2}} s_{ij}^*\delta_{d_i-d_j})\in
H^0(Z,\R)$$ and for each
$h=(0,-h_2,\ldots,-h_{N-1},h_2+\cdots+h_{N-1})\in
[({\mathbb{Z}}/{m{\mathbb{Z}}})^g ]^{N-2}\subset B(\R)$ we denote
by $\delta_h$ the corresponding global section of $\R$.

With these notations one has:

\begin{prop}\label{p:1.7}
\

\begin{enumerate}
\item
$$\begin{aligned}
\varphi^*_N(\delta_d)&=\theta[d_1](x_1+\cdots+x_N)\prod_{j\geq2}\theta[d_j](x_1-x_j)\prod_{i>j
\atop j\geq 2}\theta[d_i-d_j](x_i-x_j)   \notag \\
&=\lambda\sum_{\displaystyle b_i \in B(L^{\otimes N}) \atop
{b_1+...+b_N=d_1 \atop {b_1-b_2=d_2 \atop {...\atop
b_1-b_N=d_N}}}}
\theta[b_1](x_1)\theta[b_2](x_2)\cdots\theta[b_N](x_N) \notag
\end{aligned}$$

$\theta[b_i](x_i)$ being the global section of $L^{\otimes N}$
defined by $\delta_{b_i}$ (in the i-th component of $X^N$) and
$\theta[d_i](z)$ the global section of $L$ defined by
$\delta_{d_i}$.

\item
    {\small
    $$\begin{aligned}
    \varphi^*_N(\delta_h)\,=\,&\theta_h(x_1+\cdots+x_N,x_1-x_2,\ldots,x_1-x_N)
    \notag
    \\
    \,=\,&\lambda\sum_{\displaystyle b_i \in B(L^{\otimes N})
    \atop {b_1+...+b_N=0 \atop {b_1-b_2=-h_2 \atop {...\atop
    {b_1-b_{N-1}=-h_{N-1}\atop
    b_1-b_N=h_2+...+h_N}}}}}
    \theta[b_1](x_1)\theta[b_2](x_2)\cdots\theta[b_N](x_N)
    \notag \\
    \,=\,&\lambda\sum_{(b_1,\ldots,b_N)}
    \left(\theta[b_1](x_1)\theta[b_2+h_2](x_2)\cdots\right.
    \\
    &\left.\cdots\theta[b_{N_1}+h_{N-1}](x_{N-1})\theta[b_N-h_2-\cdots-h_{N-1}](x_N)\right)
    \notag \end{aligned}$$}

\end{enumerate}

where
$(b_1,\ldots,b_N)\in\mathrm{ker}(\xymatrix{B(\M)\ar[r]^{\varphi^*_N}&
B(\R))}$ and in both formulae $\lambda$ is a constant independent
of $d$ and $h$.
\end{prop}

\begin{demo}
This follows easily from Theorem~\ref{t:1.6} and the description
of $K(\R)$.
\end{demo}

\begin{rem}
In the case $(m,N)=1$, a general element of $B(\R)$ takes the
form:
$$d=(d_1,d_2-h_2,\ldots,d_{N-1}-h_{N-1},d_N+h_2+\cdots+h_{N-1})$$
where $(d_1,\ldots,d_N)\in B(L)^N$ and $(h_2,\ldots,d_{N-1})\in
[({\mathbb{Z}}/{N{\mathbb{Z}}})^g ]^{N-2}$, and the general
addition formula is:

$$\begin{aligned}
\varphi^*_N(\delta_h)&=\theta_h(x_1+\cdots+x_N,x_1-x_2,\ldots,x_1-x_N)
\notag \\ &=\lambda\sum_{\displaystyle b_i \in B(L^{\otimes N})
\atop {b_1+...+b_N=d_1 \atop {b_1-b_2=d_2-h_2 \atop {...\atop
{b_1-b_{N-1}=d_{N-1}-h_{N-1}\atop
b_1-b_N=d_N+h_2+...+h_N}}}}}\theta[b_1](x_1)\theta[b_2](x_2)\cdots\theta[b_N](x_N)
\notag \end{aligned}$$
\end{rem}

\begin{rem}
We have explicitly  computed the homomorphism of vector spaces
$\xymatrix{\varphi^*_N\colon H^0(Z,\R)\ar[r]  &   H^0(Z,\M)}$. If
we wish to compute:
 $$\xymatrix{\xi^*_N\colon
H^0(X^r,{\mathcal{L}})\ar[r]  &   H^0(Z,\M)}$$ let us note that we
have the commutative diagram:

$$\xymatrix{H^0(X^r,{\mathcal{L}})\ar[r]^{\xi^*_N}\ar[d]_{i^*}    &
H^0(Z,\M)   \\
            H^0(Y,{\mathcal{L}}_{|_Y})\simeq H^0(Z,\R)\ar[ur]_{\phi^*_N\equiv\varphi^*_N}  &   }$$
and we have: $$K({\mathcal{L}})\simeq X^r_m,\ \ \
K({\mathcal{L}})\cap Y\subseteq K({\mathcal{L}}_{|_Y})\simeq
K(\R)$$
$$K({\mathcal{L}})\cap Y\simeq X_m\times\overset{N}\cdots \times
X_m$$

From these identities one can easily prove that the vector
subspace $\xi_N^*H^0(X^r,{\mathcal{L}})\subseteq H^0(Z,\M)$ can be
identified with the subspace generated by the global sections
$\{\varphi_N^*(\delta_d)\}$ defined in 2.6.(a) .
\end{rem}

\section{Vector spaces of higher order odd theta functions}\label{s:2}

We shall apply the results of the first section to compute some
vector spaces of theta functions which are relevant in the study
of the fractional quantum Hall effect, ( for a similar discussion
for elliptic curves see \cite{5}) .

Following the same notations as in the previous section, let us
set an invertible sheaf $L_m={\mathcal{O}}_X(m\Theta)$ on the
principally polarized Abelian variety $(X,\Theta)$ of dimension
$g$.

Let us assume that $k=mN$ and let $L_k$ be the invertible sheaf
${\mathcal{O}}_X(k\Theta)$; on $X\times\overset{r}\cdots \times
X=Z$ we consider the invertible sheaf:
    $$
    \M=p_1^*L_k\otimes\cdots P_N^*L_k\simeq p_1^*L_k^{\otimes N}
    \otimes\cdots P_N^*L_k^{\otimes N}
    $$

Let us define the vector subspace $E_k(N)\subset H^0(Z,\M)$ by the
following conditions: $$s\in E_k(N)\Longleftrightarrow
\text{\parbox{9.5cm}{$s$ is invariant with respect to the action
of the $N$-torsion subgroup $\Delta(X_N)\subset Z$ and is odd with
respect to the permutations acting on
$H^0(Z,\M)=H^0(X,L_k)\otimes\cdots\otimes H^0(X,L_k)$}}$$

Let us set $V_m=H^0(X,L_m)$ and $V_k=H^0(X,L_k)$. By the very
definition, one has that: $$E_k(N)=\bigwedge^N V_k \cap
\mathrm{Im}\varphi^*_N\subset V_k\otimes\overset{N}\cdots \otimes
V_k$$

where $\xymatrix{\varphi^*_N\colon H^0(Z,\R)\ar[r]  &
H^0(Z,\M)=V_k\otimes\overset{N}\cdots \otimes V_k}$ is the
addition homomorphism defined in the last section.

Note that the factorization $\varphi^*_N=\pi_{1\ldots N}\circ
\xi_N$ implies that: $$E_k^0(N)=\bigwedge^N V_k \cap
\mathrm{Im}\xi^*_N\subseteq E_k(N)\subseteq V_k^{\otimes N}$$

Let $E_i^{\pm}\subset H^0(Z,\R)$ be the subspaces of eigenvectors
of the automorphism on $H^0(Z,\R)$ induced by
$\xymatrix{\sigma_i\colon X^N\ar[r] &   X^N},\
\sigma_i(x_1,\ldots,x_N)=(x_1,\ldots,-x_i,\ldots,x_N)$.

\begin{prop}\label{p:2.1}
There exists a natural isomorphism: $$E_k(N)\simeq\varphi^*_N
H^0(Z,\R)_-$$ $H^0(Z,\R)_-$ being the vector subspace of
$H^0(Z,\R)$ defined as the intersection of the vector subspaces
$E_i^-$ with $i>1$.
\end{prop}

\begin{demo}
This is easy from the equality $E_k(N)=\bigwedge^N V_k\cap
\mathrm{Im}\varphi^*_N$.
\end{demo}

We can give a more explicit description of the subspace
$E^0_k(N)$.

From the Remark 2.3 , it follows that $\mathrm{Im}\xi^*_N$ is the
vector subspace of $H^0(Z,\R)$ described in
Proposition~\ref{p:1.7}.

Let $V^{\pm}_m$ be the subspaces of eigenvectors of $V_m$ with
respect to the action of the involution
 $\xymatrix{[-1]_X\colon X\ar[r]    &   X}\ \ ([-1]_X(x)=-x)$. Then, we have:

\begin{prop}\label{p:2.2}
$$E^0_k(N)=\varphi^*_N(V_m\otimes V_m^- \otimes\overset{N-1}\cdots
\otimes V_m^-)$$
\end{prop}

\begin{demo}
One has only to observe that $\mathrm{Im}\xi^*_N$ is naturally
identified with $V_m\otimes\overset{N}\cdots \otimes V_m$.
\end{demo}

In our interpretation of the FQHE, the vector subspace $E^0_k(N)$
is the space of wave functions of a system of $N$ electrons.

\section{Poincar\'e bundles and Fourier-Mukai transforms}\label{s:3}

Let $(X,\Theta)$ be a p.p.a.v. of dimension $g$ and $\hat{X}$ its
dual Abelian variety. Let $\poin$ be a Poincar\'e bundle on
$X\times\dua$; $\poin$ is the line bundle on $X\times\dua$ given
by the universal property of $\dua$.

Given an invertible sheaf $L_m\simeq{\mathcal{O}}_X(m\Theta)$ on
$X$ (with $m>0$), we can construct the invertible sheaf on
$X\times\dua$: $${\mathcal{L}}_m=\pi_X^*L_m\otimes\poin$$ where
$\xymatrix{\pi_X\colon X\times\dua\ar[r]  &   X}$ and
$\xymatrix{\pi_{\dua}\colon X\times\dua\ar[r]  &   \dua}$ are the
natural projections.

The Fourier-Mukai transform of $L_m$ is ( see \cite{6} and
\cite{7} for details):
$$S(L_m)={\pi_{\dua}}_*\left(\pi_X^*L\otimes\poin\right)={\pi_{\dua}}_*{\mathcal{L}}_m$$

It is well known that $S(L_m)$ is a rank $m^g$ vector bundle on
$\dua$.

We can interpret $\Lm$ as the family of line bundles over $X$,
parametrized by $\dua$, which are algebraically equivalent to
$L_m$.

If we wish to generalize the results of Section.1 to the case of a
``variable line bundle" $L_m$, we must perform the base change
$\xymatrix{X\times\dua\ar[r]    &   \dua}$ and replace $L_m$ by
$\Lm$.

We can then define on $Z\times\dua$ the following line bundles:

    $$\begin{aligned}
    &\hm=\overline{M}^*\Lm\otimes\left(\bigotimes_{i<j}
    \overline{s}_{ij}^*\Lm\right) \notag
    \\
    &\hr=
    \left(
    \overline{p}_1^*\Lm \otimes\cdot\cdot\cdot
    \otimes\overline{p}_N^*\Lm\right)
    \otimes\left(\bigotimes_{i>j\atop j\geq 2}
    \overline{s}_{ij}^*\Lm \right) \notag
    \end{aligned}$$

where $\overline{M}$ and $\overline{s}_{ij}$ are the morphisms
$\xymatrix{Z\times\dua\ar[r] &   X\times\dua}$ defined by:
$\overline{M}=M\times \mathrm{Id}_{\dua},\ \
\overline{s}_{ij}=s_{ij}\times\mathrm{Id}_{\dua}$ and
$\xymatrix{\overline{p}_i\colon Z\times\dua\ar[r]   &
X\times\dua}$ are the natural projections.

Defining $\xymatrix{\overline{\varphi}_N\colon Z\times\dua\ar[r]
&   Z\times\dua}$ as
$\overline{\varphi}_N=\varphi_N\times\mathrm{Id}_{\dua}$ we have
that: $$\overline{\varphi}_N^*\hr\simeq\hm$$

and Corollary 2.2 implies that:
$$\overline{\varphi}_N^*\hr\simeq\hm\simeq
\overline{p}_1^*\Lm^{\otimes
N}\otimes\cdots\overline{p}_N^*\Lm^{\otimes
N}\otimes\pi_{\dua}^*F$$ for some invertible sheaf $F$ on $\dua$.

Bearing in mind the applications to the FQHE, we are mainly
interested in the bundles:
$$W_N(L_m)={\pi_{\dua}}_*(M^*\Lm)={\pi_{\dua}}_*\left(\overline{M}^*\left(\pi_X^*L_m\otimes\poin\right)\right)$$
which describe the dynamics of the center of mass.

Our main result on the structure of $W_N(L_m)$ is as follows:

\begin{teo}\label{t:3.1}

For every $N>0$ and $m>0$, $W_N(L_m)$  are vector bundles over
$\dua$ of rank $m^g$. These vector bundles are semistable with
respect to the principal polarization $\hat{\Theta}$ induced by
$\Theta$ on $\dua$. Moreover, for every $N\geq 2$, there exist
natural isomorphisms $\xymatrix{W_N(L_m)\ar[r]^{\sim}   &
W_{N-1}(L_m)}$.
\end{teo}

\begin{demo}
Proof of the existence of isomorphisms
$\xymatrix{W_N(L_m)\ar[r]^{\sim}   &   W_{N-1}(L_m)}$ is the same
as the proof given
 in the case of elliptic curves.

Therefore, the proof of the theorem is reduced to the case of
$W_1(L_m)$ which is precisely the Fourier-Mukai transform of
$L_m$, which is well known to be a vector bundle of rank $m^g$
(for $m>0$).

We only have to prove the semi-stability of $W_1(L_m)$ with
respect to $\hat{\Theta}$.

Let us compute the slope of $W_1(L_m)$: we consider the isogeny
$\varphi_{L_m}:X \to \dua$ of degree $m^{2g}$ defined by:
$$\varphi_{L_m}=T_x^*L_m\otimes L_m^{\otimes -1}$$
$\xymatrix{T_x\colon X\ar[r]    &   X}$ being the translation by
$x$. It is known (\cite{8}) that:
$$\varphi_{L_m}^*W_1(L_m)\simeq H^0(X,L_m)\otimes L_m^{\otimes
-1}$$

Let us set ${\mathcal{O}}_X(D)=\mathrm{det}W_1(L_m)$ ; one has:
$$\varphi^*(D\cdot\hat{\Theta}^{g-1})=\mathrm{deg}(\varphi)\cdot\mathrm{deg}(D)=m^{2g}\cdot\mathrm{deg}(D)$$
and
$$\varphi^*(D\cdot\hat{\Theta}^{g-1})=\varphi^*D\cdot(\varphi^*\hat{\Theta})^{g-1}=(-m^{g+1}\Theta)\cdot(m^2\Theta)^{g-1}=-m^{3g-1}g!$$
Then, $\mathrm{deg}(D)=-m^{g-1}g!$.

$$\mu(W_1(L_m))=\frac{\mathrm{deg}W_1(L_m)}{\mathrm{rk}W_1(L_m)} =
-\frac{g!}{m}$$

Let us recall that from the computations of \cite{9}  one  easily
deduces that given an invertible sheaf $\mathcal{M}$ on $\dua$,
one has that $c_1({\mathcal{M}})\cdot{\hat{\Theta}}^{g-1}=g!\cdot
c$ for some integer $c$. Thus, in the definition of semi-stability
on $\dua$, with respect to the polarization $\hat{\Theta}$, we can
replace the degree
$c_1({{\mathcal{M}}})\cdot{\hat{\Theta}}^{g-1}$, of an invertible
sheaf ${\mathcal{M}}$, by the reduced degree:
$$r\mathrm{deg}({\mathcal{M}})=\frac{c_1({\mathcal{M}})\cdot{\hat{\Theta}}^{g-1}}{g!}$$
and the reduced slope:
$$\mu_r({\mathcal{M}})=\frac{c_1({\mathcal{M}})\cdot{\hat{\Theta}}^{g-1}}{g!\cdot\mathrm{rk}{\mathcal{M}}}$$

Let $F\subseteq W_1(L_m)$ be a subbundle of rank $r<m^g$ and
reduced degree $r\mathrm{deg}(F)=d$. One has to show that:
$$\mu_r(F)=\frac{d}{r}\leq\mu_r\left(W_1(L_m)\right)=-\frac{1}{m}$$

But it is known that to prove the semi-stability condition for
$W_1(L_m)$ it suffices to prove that it is satisfied by the
subbundles of $\mathrm{rk}=1$; that is, we can assume that $r=1$.
In this case, the inequality      is equivalent to $d<0$.

Let us take the pullback of $F\subseteq W_1(L_m)$ with respect to
the isogeny $\varphi_{L_m}$:
$$\varphi^*_{L_m}F\subseteq{\varphi_{L_m}}_*W_1(L_m)\simeq
H^0(X,L_m)\otimes L_m^{\otimes -1}$$ and
$r\mathrm{deg}\varphi_{L_m}^*F=m^{2g}d\leq
r\mathrm{deg}L_m^{\otimes -1}=-m<0$. Then, one has that $d<0$.

\end{demo}

\section{Fractional quantum Hall states in multi-layer two-dimensional electron systems}

For applications to the FQH effect, we shall apply the theory
developed in previous sections to the following situation:

Let us consider the forms
$E={\mathbb{C}}/{{\mathbb{Z}}\oplus\tau{\mathbb{Z}}}$ defined by
$\tau\in {\mathbb{H}}_1$ (upper half-plane) and let us denote by
$e\in E$ the origin of the group law of $E$. The natural
polarization on $E$ is given by the invertible sheaf
${\mathcal{O}}_E(e)$.

For any positive integer $g\in{\mathbb{Z}}$, let us denote by
$X_g$ the abelian variety: $$X_g=E\times\overset{g}\cdots \times
E$$

Let $\xymatrix{X_g\ar[r]^{q_i}  &   E}$ be the natural projection
into the i-th factor. One can define a principal polarization,
$\Theta$, on $X_g$ as follows:
$${\mathcal{O}}_X(\Theta)=\bigotimes_{i=1}^gq_i^*{\mathcal{O}}_E(e)$$

Let $K$ be a symmetric, positive, integer-valued $g\times g$
matrix. This matrix defines an isogeny: $$\xymatrix{K\colon
E^g=X\ar[r] &   E^g=X}$$

One can define a line bundle $L_k$ on $X$ by:
$$L_k=K^*{\mathcal{O}}_X(\Theta)$$

We can apply  the results of  Sections 1 and 2 to this sheaf.

Let $N>0$ be an integer number, $r=\frac{N(N-1)}{2}+1$, and
$\xi_N,\ \varphi_N$ the morphisms defined in Section.1

$$\xymatrix{\xi_N\colon X\times\overset{N}\cdots\times X\simeq
E^{gN}\ar[r] &   X\times\overset{r}\cdots \times X}$$

$$\xymatrix{\varphi_N\colon X\times\overset{N}\cdots \times
X\ar[r] &   X\times\overset{N}\cdots \times X}$$

On $Z=X^N$, one has the sheaf:
    $$
    \R=(p_1^*L_K\otimes\cdots\otimes p_N^*L_K)\otimes
    \left(\bigotimes_{i>j \atop j\geq 2} s_{ij}^*L_K\right)
    $$
and isomorphisms: $$\xi^*_N\left(\bigotimes_{i=1}^r
p_i^*L_K\right)\simeq\varphi^*_NR_N\simeq\bigotimes_{i=1}^N
p_i^*L_K^{\otimes N}\simeq\M(K)$$

Analogously to Section.2  , for each matrix $K$ we can define the
vector subspace:
 $$E_K(N)\subset H^0(Z,\M(K))$$ which will be
identified with the Hilbert space of our problem:

$$s\in E_K(N)\Longleftrightarrow \text{\parbox{9.8cm}{$s$ is
invariant with respect to the action of the subgroup
$\Delta(X_N)\subset Z$ and is odd with respect to the permutations
acting on $H^0(Z,\M(K))=H^0(X,L_K^{\otimes N})\otimes\cdots\otimes
H^0(X,L_K^{\otimes N})$}}$$

Also one has that: $$E_K(N)=\bigwedge^N H^0(X,L_K^{\otimes N})\cap
\mathrm{Im}\varphi^*_N$$

Analogously to Section.2 we can also define the subspace
$E^0_K(N)=\bigwedge^N H^0(X,L_K^{\otimes N})\cap
\mathrm{Im}\xi^*_k$

Let us denote a point of $X^N$ by $(x_1,\ldots,x_N)$ and
$x_i=(t_1^i,\ldots,t_g^i)\in E^g=X$.

The explicit computations can be performed along the lines of
(\cite{2}) and (\cite{4}).

Note that the kernel of the isogeny $\xymatrix{K\colon X\ar[r]
&   X}$ can be identified with the finite subgroup:
$$X_K\simeq{\mathbb{Z}}^g/{K{\mathbb{Z}}^g}\times
\hat{{\mathbb{Z}}^g/{K{\mathbb{Z}}^g}}$$

The order of this group is $|X_K|=|\mathrm{det}K|^2$ and
$H^0(X,L_K)$ is a ${\mathbb{C}}$-vector space of dimension
$|\mathrm{det}K|$. Obviously one has: $$K(L_K)=X_K\subset
K(L_K^{\otimes N})$$ $$N\cdot K(L_K^{\otimes N})=K(L_K)$$

Let us set $V=H^0(X,L_K)$ and $V_K=H^0(X,L_K^{\otimes N})$. One
has the analogous results of those proved in Sections 2 and 3 and
: $$\begin{aligned} &E_K(N)=\varphi_N^*H^0(Z,\R)_-  \notag \\
&E^0_K(N)=\varphi^*_N(V\otimes V^-\otimes\overset{N-1}\dots
\otimes V_-) \notag \end{aligned}$$

Moreover, given $d=(d_1,\ldots,d_N)\in N\cdot
B(L_K)^N=[{\mathbb{Z}}^g/{K{\mathbb{Z}}^g}]^N\subseteq B(\R)$, let
us denote by $\delta_d$ the element
$$\delta_d=\delta_{d_1}\otimes\cdots\otimes\delta_{d_N}\otimes\left(\bigotimes_{i>j
\atop j\geq 2} s_{ij}^*\delta_{d_i-d_j}\right)\in H^0(Z,\R)$$

It follows that the vector subspace $E^0_K(N)$ is generated by the
sections $\varphi_N^*(\delta_d)$ and one has the identity:

$$\begin{aligned}
\varphi_N^*(\delta_d)&=\theta[d_1](x_1+\cdots+x_N)\prod_{j\geq2}\theta[d_j](x_1-x_j)\prod_{i>j
\atop j\geq 2}\theta[d_i-d_j](x_i-x_j)  \notag  \\
&=\lambda\sum_{\displaystyle b_1+...+b_N=d_1 \atop {b_1-b_2=d_2
\atop {...\atop
b_1-b_N=d_N}}}\theta[b_1](x_1)\theta[b_2](x_2)\cdots\theta[b_N](x_N)
\label{5.1}
\end{aligned}$$
where $(x_1,\ldots,x_N)\in X\times\overset{N}\cdots \times
X=E^{gN}$ that is, $x_i=(z_{i1},\ldots,z_{ig})\in E^g$.

Observe that: $${\mathbb{Z}}^g/{K{\mathbb{Z}}^g}\simeq
{\mathbb{Z}}/{n_1{\mathbb{Z}}}\oplus\cdots\oplus
{\mathbb{Z}}/{n_g{\mathbb{Z}}}$$ for some integers
$n_1,\ldots,n_g$ such that $\mathrm{det}K=n_1\cdots n_g$.

Then, in the above statements $d_1,\ldots,d_N$ are elements of the
group ${\mathbb{Z}}/{n_1{\mathbb{Z}}}\oplus\cdots\oplus
{\mathbb{Z}}/{n_g{\mathbb{Z}}}$ (once one has fixed the
corresponding theta-structures).

\section{Filling factors and Hall conductivity}

In a multi-layer many-electron system where the fractional quantum
Hall effect is observed, the ground state is a quantum fluid with
several possible topological orders; see \cite{10} . The different
phases are characterized by the $g\times g$ matrix: $$K=\left(
\begin{array}{cccc}
                2p+1   &   2p     &   \cdots    &   2p   \\
                2p     &   2p+1   &   \cdots    &   2p   \\
                \vdots &   \vdots &   \ddots    &   \vdots \\
                2p     &   2p     &   \cdots    &   2p+1    \end{array} \right),$$
where $p$ is an integer greater than zero and $g$ is the number of
layers.

The ground state wave function
 $$\tilde{\psi}=\prod_{i,j=1 \atop i<j}^N[\prod_{a=1}^g(z_i^a-z_j^a)^{2p+1}\prod_{a<b}(z_i^a-z_j^b)^{2p}]\textrm{exp}[-\sum_{a=1}^g\sum_{i=1}^N|z_i^a|^2]$$
is the generalization of the Laughlin state to the case in which
each layer is isomorphic to ${\mathbb{C}}$; here, $z_i^a$ is the
$i$-th particle position in the $a$-th layer, and we assume that
there are $N$ particles per layer, so that the total number of
particles is $N_T=gN$.

We focus on this problem when each electron moves on a torus; the
one-particle configuration space is the elliptic curve
$E={\mathbb{C}}/{{\mathbb{Z}}+\tau{\mathbb{Z}}}$ of the previous
Sections. The modular parameter $L_2e^{i\theta}/L_1$ encodes the
periodicities of the basic lattice, which is the same for every
layer. A constant magnetic field $B$ allows for a well behaved
quantum system, compatible with the lattice and the order ``meant"
by the matrix $K$, if and only if: $$K_1=\left(
\begin{array}{cccc}
                (2p+1)N   &   2pN    &   \cdots    &   2pN   \\
                2p        &   2p+1   &   \cdots    &   2p   \\
                \vdots    &   \vdots &   \ddots    &   \vdots \\
                2p        &   2p     &   \cdots    &   2p+1    \end{array} \right),
\ \ \frac{eB}{\hbar
c}L_1^2=\frac{2\pi|\mathrm{det}K_1|}{\mathrm{Im}\tau}.    $$

Here, $e,\hbar$ and $c$ are respectively the electron charge, the
Planck constant and the speed of light in vacuum. The quantum
space of one-particle states is the space of sections of the line
bundle $L_{K_1}=K_1^*\theta_{X_g}(\Theta)$ and the first Landau
level corresponds to the sub-space of holomorphic sections
$H^0(X_g,L_{K_1})$.

There is a many-electron wave function proposed by Haldane and
Rezayi as the ground state for the quantum Hall fluid in a
periodic lattice, see \cite {11} and \cite {12}. Both the HR wave
function and its generalization to a multi-layer are of Laughlin
type and the framework for the mathematical understanding of such
complex quantum states is provided by the developments set forth
before in this paper. We start by noticing that the isomorphism
established at the end of Section .4 now reads:
$${\mathbb{Z}}^g/{K{\mathbb{Z}}^g}\simeq{\mathbb{Z}}/{(2gp+1){\mathbb{Z}}}\oplus
1\oplus 1\oplus\cdots\oplus 1$$ i.e.
$n_1=(2gp+1),n_2=n_3=\cdots=n_g=1$ because these are the
eigenvalues of the $K$ matrix.

The center-of-mass dynamics and the relative motion of each pair
of particles produce  contributions that factorize in the ground
state wave function. In a basis in $X_g$ in which $K$ is diagonal:
\begin{enumerate}
\item

The center-of-mass wave function is a Theta function of $g$
variables that we write following the conventions of Reference
\cite {13}  in order to translate the developments of the previous
Sections to the notation used in the physics literature:
$$F_{CM}(\vec{X})=\Theta \left[
\begin{array}{c}
        d_1K_D^{-1}\vec{e}_1   \\
        \vec{0}                \end{array}
\right](K_D\vec{X}\ |\ K_D\tau) $$
$$\vec{X}=\vec{x_1}+\vec{x_2}+\cdots+\vec{x_N}  $$ $\vec{X}$ is
the CM coordinate, $K_D$ is a diagonal matrix such that
$\mathrm{det}K_D=\mathrm{det}K$ (we have chosen
$K_{D_{11}}=2gp+1$) and the vector of $g$ components $\vec{e}_1$
is $(1,0,\ldots,0)$.

This expression for the center-of-mass wave function is exactly
the same as $\theta[d_1](x_1+x_2+\cdots+x_N)$ in the previous
Section and, undoing the diagonalization, one obtains:
$$F_{CM}(\vec{Z})=\Theta \left[
\begin{array}{c}
        K^{-1}\vec{\alpha}   \\
        \vec{0}                \end{array}
\right](K\vec{Z}\ |\ K\tau) $$ where
$\vec{Z}=\vec{z_1}+\vec{z_2}+\cdots+\vec{z_N}$ is the CM
coordinate in a basis of $X_g$ where $K$ is not diagonal, and
$\vec{\alpha}\in{\mathbb{Z}}^g/{K{\mathbb{Z}}^g}$. This is the
form in which it appears in the physics literature.

\item

The factor in the ground state wave function due to relative
motion has the form: if $\vec{x}_{ij}=\vec{x}_{i}-\vec{x}_{j}$,
$$F_{r}(\vec{x}_{ij})=\prod_{i<j}\Theta_- \left[
\begin{array}{c}
        d_{ij}^-K_D^{-1}\vec{e}_1   \\
        \vec{0}                     \end{array}
\right](K_D\vec{x}_{ij}\ |\ K_D\tau) $$ $$d_{ij}^-=d_i^--d_j^-,
i\geq2,d_{ij}=d_j^- , i=1, d_{ij}^-=1,2,\ldots,gp$$

Fermi statistics requires the use of anti-symmetric functions in
$\vec{x}_{ij}\mapsto -\vec{x}_{ij}$: {\small $${\displaystyle
\Theta_- \left[
\begin{array}{c}
        d_{ij}^-K_D^{-1}\vec{e}_1   \\
        \vec{0}                \end{array}
\right](K_D\vec{x}_{ij}\ |\ K_D\tau) \atop = \displaystyle
\frac{1}{2}\left(\Theta \left[
\begin{array}{c}
        d_{ij}^-K_D^{-1}\vec{e}_1   \\
        \vec{0}                \end{array}
\right](K_D\vec{x}_{ij}\ |\ K_D\tau)- \Theta \left[
\begin{array}{c}
        -d_{ij}^-K_D^{-1}\vec{e}_1   \\
        \vec{0}                \end{array}
\right](K_D\vec{x}_{ij}\ |\ K_D\tau)\right)}$$}

Nevertheless, the ground state wave function
$$\psi=F_{CM}(\vec{X})F_{r}(\vec{x}_{ij})\mathrm{exp}\{-\frac{1}{4}\sum_{i}(\mathrm{Im}\vec{x}_{i})(\mathrm{Im}\vec{x}_{i})\},$$
apart from the non-analytic exponential factor, consists of terms
of the form of the left-hand member of formula ( \ref{5.1} ) in
Section \S.5 .

Therefore, $\psi$ can also be expressed as a product of Theta
functions in the $\vec{x}_{i}$ variables  with characteristics:
$$b_i\in{\mathbb{Z}}/{(2gp+1){\mathbb{Z}}}\oplus 1\oplus
1\oplus\cdots\oplus 1$$ related to the $K_D$ matrix.

In the physics of the quantum Hall effect, the concept of the
filling factor plays a central r\^ole; if the magnetic field is
strong enough to provide more states in the first Landau level
than electrons, it is defined as: $$f=\frac{\text{number of
particles}}{\text{number of states in the first $LL$}},$$ and the
Hall conductivity is studied as a function of $f$.

If the number of states in the first $LL$ is a finite number,
$\mathrm{dim}H^0(X_g,L_{K_1})=\mathrm{det}(K_1)$ in our case, then
$f$ is: $$f_{HR}=N_T/{\mathrm{det}K_1}=\frac{g}{2gp+1}.$$

Different integers $g$, and hence different values of $f_{HR}$,
give rise to a hierarchy of  experimentally observed topological
orders: associated with each $f$ of this form there are quantum
fluids that arise as ground states of the fractional quantum Hall
effect without periodic boundary conditions.

What we have shown by proving the generalized addition formulae
for Abelian varieties is that the fractional quantum Hall states
in multi-layer two-dimensional electron systems are compatible
with periodic lattices. Only the existence of such addition
formulae makes it possible to claim that the generalized
Haldane-Rezayi wave function implies $f_{HR}=\frac{g}{2gp+1}.$

In fact, a further development remains to be made in order to make
contact with the HR ground state. We remark that there is a linear
combination such that: $${\displaystyle \sum_{d_{ij}^-=1}^{gp}
c[d_{ij}^-] \Theta_- \left[
\begin{array}{c}
        d_{ij}^-K_D^{-1}\vec{e}_1   \\
        \vec{0}                \end{array}
\right](K_D\vec{X}_{ij}\ |\ K_D\tau)\atop
=
\Theta^{2gp+1} \left[
\begin{array}{c}
        1/2   \\
        1/2 \end{array}
\right](x_i^1-x_j^1\ |\ \tau) \prod_{a=2}^g \Theta \left[
\begin{array}{c}
        1/2   \\
        1/2 \end{array}
\right](x_i^a-x_j^a\ |\ \tau)}, $$ appearing in the right-hand
member odd Theta functions of one variable. Undoing the
diagonalization of $K$, one easily checks that, $${\displaystyle
\prod_{a=1}^g \Theta^{2p+1} \left[
\begin{array}{c}
        1/2   \\
        1/2 \end{array}
\right](z_i^a-z_j^a\ |\ \tau) \prod_{a<b} \Theta^{2p} \left[
\begin{array}{c}
        1/2   \\
        1/2 \end{array}
\right](z_i^a-z_j^a\ |\ \tau) \atop \simeq \Theta^{2gp+1} \left[
\begin{array}{c}
        1/2   \\
        1/2 \end{array}
\right](x_i^1-x_j^1\ |\ \tau) \prod_{a=2}^g \Theta \left[
\begin{array}{c}
        1/2   \\
        1/2 \end{array}
\right](x_i^a-x_j^a\ |\ \tau)}$$ in such a way that the HR wave
function can be traced back to the above $\psi$.

The generalized addition formulae are valid for any Abelian
variety
$X_g=\frac{{\mathcal{C}}^g}{{\mathcal{Z}}^g\oplus\Omega{\mathcal{Z}}^g}$,
with $\Omega$ a matrix in the Siegel upper half-space of rank $g$
in ${\mathbb{H}}_g$. In the application to the quantum Hall
effect, we have restricted ourselves to the case $X_g=E^g$, i.e.
$\Omega=\tau I_{g\times g}$. There is no difficulty in extending
the analysis to any $\Omega\in{\mathbb{H}}_g$ that physically
corresponds to taking into account different periodicities for
different layers and a tunnel effect of weak amplitude between
layers, a situation also considered by condensed matter
physicists, see \cite {14}. It is also convenient to make a brief
comment on the second type of addition formulas;
Proposition~\ref{p:1.7} of Section .1  , from a physical point of
view. Mathematically, the origin of such addition formulas is the
freedom of choosing the isogeny $\xymatrix{\varphi_N\colon Z\ar[r]
&   Z}$: there are different projections from $X_g\times
\overset{r} \cdots\times X_g$ to
$Z=X_g\times\overset{N}\cdots\times X_g$ ($r=\frac{N(N-1)}{2}+1$).
Another choice of $\pi_{1\ldots N}$, for instance, would lead one
define $$\varphi_N
'(x_1,\ldots,x_N)=(x_1+x_2+\cdots+x_N,x_2-x_1,x_2-x_3,\ldots,x_2-x_N),$$
i.e. it would singularize relative coordinates with respect to the
second particle. In quantum mechanics particles are
indistinguishable and thus this possibility is physically
equivalent to choosing $\varphi_N$ based on the first particle
coordinate. For this reason the wave functions invariant under the
second sub-group of $K(\R)$ do not enter in physical arguments,
and the ordering $i<j$ is chosen as the most natural one.

Further  knowledge of the implications of the nature of the HR
ground state wave function can be obtained by means of a gedanken
experiment, see \cite {15}  : magnetic fluxes are induced by two
solenoids per layer connected to the Hall device in such a way
that they are compatible with the electrons if:
$$\mathrm{Re}\vec{\phi}\in[\vec{0},\frac{\hbar c}{e}\vec{u}],\ \ \
\mathrm{Im}\vec{\phi}\in [\vec{0},\frac{\hbar c}{e}\vec{u}]$$
according to the Aharanov-Bohm effect. Here, $\vec{\phi}$ is a
complex $g$ vector which encodes the solenoid fluxes and
$\vec{u}=(1,1,\ldots,1)$ is a real constant $g$ vector. The
generalized HR states are modified to,
$$\psi_{d_1}[\vec{\phi}]=F_{CM}^{d_1}[\vec{\phi};\vec{X}]F_r[\vec{X}_{ij}]\mathrm{exp}\{-\frac{1}{4}\sum_{i}[(\mathrm{Im}\vec{X}_{i})^t\mathrm{Im}\vec{X}_{i}]\}$$
$$F^{d_1}_{CM}[\vec{\phi};\vec{X}]= \Theta \left[
\begin{array}{c}
        d_1 K_D^{-1}\vec{e}_1+\vec{\phi}_1   \\
        \vec{\phi}_2                \end{array}
\right](K_D\vec{X}_{ij}\ |\ K_D\tau)$$ where
$\vec{\phi}_1=\frac{e}{\hbar c}\mathrm{Re}\vec{\phi}$ and
$\vec{\phi}_2=\frac{e}{\hbar c}\mathrm{Im}\vec{\phi}$. The
relative motion is not affected but the contribution of the
center-of-mass dynamics to the ground state is modified by
including the solenoid fluxes as characteristics of the Theta
function.

Mathematically, one must interpret $\vec{\phi}$ as points in the
Jacobian $\dua_g$ of $X_g$ and we proceed to identify the bundle
where $\psi^{d_1}[\vec{\phi}]$ is defined as a section, using the
developments of Section .3  . In fact, only the replacement of
$L_m$ by $L_K$ is necessary. We thus start  by constructing the
invertible sheaf: $${\mathcal{L}}_K=\pi_X^*L_K\otimes\poin,$$ a
family of line bundles over $X$ parametrized by $\dua$, and
defining the Fourier-Mukai transform of $L_K$:
$$S(L_K)={\pi_{\dua}}_*(\pi_X^*
L_K\otimes\poin)={\pi_{\dua}}_*{\mathcal{L}}_K.$$

$S(L_K)$ is a vector bundle over $\dua$ of rank
$(\mathrm{det}K)^g$ whose fibers are vector spaces of dimension
$(\mathrm{det}K)^g$ whose bases are provided by the basis of
$H^0(\dua,L_K)_{|_{\hat{x}_0\in\dua}}$. Taking this into account,
one easily recognizes that
$$s^{d_1}=F_{CM}^{d_1}[\vec{\psi};\vec{X}]F_r[\vec{x}_{ij}]$$ is a
holomorphic section in the bundle
$$\tilde{{\mathcal{M}}}_N^K=\overline{M}^*{\mathcal{L}}_K\otimes\left(\bigotimes_{i<j}\overline{s}_{ij}^*{\mathcal{L}}_K\right)$$
defined in perfect analogy with the bundle
$\tilde{{\mathcal{M}}}_N$ of Section .3  : one merely replaces
${\mathcal{L}}_m$ by ${\mathcal{L}}_K$.

We now focus on the center-of-mass dynamics. Taking direct image
amounts to integrate over the variables in the other factors and
we find $$S_{CM}^{d_1}=F_{CM}^{d_1}[\vec{\phi}]=\int_X
{d\mathrm{vol}_X F_{CM}^{d_1}[\vec{\phi};\vec{X}]},$$ which
determines the contribution of the solenoid fluxes to the CM
ground state wave function; this is a holomorphic section in the
Fourier-Mukai transform of the bundle $\overline{M}^*L_K$:
$$W_N(L_K)={\pi_{\dua}}_*(\overline{M}^*{\mathcal{
L}}_K)={\pi_{\dua}}_*\left(\overline{M}^*({\pi_X}^*L_K\otimes
\poin)\right).$$

From Section .3  we know that $W_N(L_K)\simeq W_{N-1}(L_K)$ and
the slope and reduced slope of $W_1(L_K)$ are given by:
$$\begin{aligned}
&\mu(W_1(L_K))=-\frac{g(\mathrm{det}K)^{g-1}g!}{(\mathrm{det}K)^g}=-\frac{gg!}{\mathrm{det}K}
\notag \\ &\mu_r(W_1(L_K))=-\frac{g}{\mathrm{det}K}       \notag
\end{aligned}$$

There is a novelty: the factor $g$ appears due to the freedom of
choosing $2gp+1$ as any of the $g$ eigenvalues of $K$.

The Hall conductivity of the system is expressed in perturbation
theory by the Kubo-Thouless formula \cite{13} :
$$\sigma_H=\frac{i}{2\pi}\cdot\frac{ge^2}{r\hbar}\sum_{d_1=1}^{r}[<\vec{\nabla}_1\psi^{d_1}\
|\ \vec{\nabla}_2\psi^{d_1}>-<\vec{\nabla}_2\psi^{d_1}\ |\
\vec{\nabla}_1\psi^{d_1}>]$$ where $r=\mathrm{det}K$,
$\vec{\nabla}_a=\frac{\partial}{\partial \vec{\phi}_a}$ and $<\ |
\ >$ defines the $L^2$-norm: $$<f\ |\ g >=\int_{X^{\otimes
N}}{d\mathrm{vol}_{X^{\otimes N}}
f^*(\vec{x}_1,\vec{x}_2,\ldots,\vec{x}_N)g(\vec{x}_1,\vec{x}_2,\ldots,\vec{x}_N)}.$$

This formula can be interpreted as follows: from the section
$\psi^{d_1}$, we obtain a connection, for any $d_1$,
$$\omega^{d_1}=-2\mathrm{Im}<\psi^{d_1}\ |\
\vec{\nabla}_2\psi^{d_1}>\cdot d\vec{\phi}_2$$ in a certain line
bundle over $\dua$. The curvature
$${\mathcal{R}}_{\omega^{d_1}}=2\pi\cdot d\vec{\phi}_1\wedge
d\vec{\phi}_2$$ is constant on $\dua$ and therefore $\sigma_H$ is
equal to its average value $<\sigma_H>$:
\begin{align}
<\sigma_H>&=\frac{i}{2\pi}\cdot\frac{ge^2}{r\hbar}\sum_{d_1=1}^{r}[<\vec{\nabla}_1\psi^{d_1}\
|\ \vec{\nabla}_2\psi^{d_1}>-<\vec{\nabla}_2\psi^{d_1}\ |\
\vec{\nabla}_1\psi^{d_1}>]       \notag \\ &=\frac{g}{2gp+1}.
\notag\end{align}

The bundle is therefore $W_1(L_K)$ and the Hall conductivity is a
topological invariant, the reduced slope of $W_1(L_N)$:
$$\sigma_H=|\mu_r(W_1(L_K))|$$

\end{enumerate}

\end{document}